\documentclass[%
superscriptaddress,
preprint,
 amsmath,amssymb,
 aps,
]{revtex4-2}
\usepackage{setspace}
\usepackage{graphicx}
\usepackage{dcolumn}
\usepackage{bm}
\usepackage[normalem]{ulem}
\usepackage{hyperref}
\usepackage{xcolor}
\hypersetup{
    colorlinks,
    linkcolor={blue},
    citecolor={blue},
    urlcolor={blue}
}

\begin{document}
\title{Non-monotonic emergence of order from chaos in turbulent thermo-acoustic fluid systems}

\author{Aswin Balaji}
 \affiliation{%
 Department of Aerospace Engineering, Indian Institute of Technology Madras, Chennai 600036, India
} 
\affiliation{%
 Centre of Excellence for Studying Critical Transitions in Complex Systems, Indian Institute of Technology Madras, Chennai 600036, India
}%
\author{Shruti Tandon}%
 
\affiliation{%
 Department of Aerospace Engineering, Indian Institute of Technology Madras, Chennai 600036, India
}%
\affiliation{%
 Centre of Excellence for Studying Critical Transitions in Complex Systems, Indian Institute of Technology Madras, Chennai 600036, India
}%
\author{Norbert Marwan}
\affiliation{%
 Potsdam Institute for Climate Impact Research (PIK), Potsdam 14473, Germany
}%
\affiliation{%
 Institute of Geosciences, University of Potsdam, Potsdam 14476, Germany
}%
\affiliation{%
Institute of Physics and Astronomy, University of Potsdam, Potsdam, 14476, Germany
}%

\author{J\"urgen Kurths}
\affiliation{%
 Potsdam Institute for Climate Impact Research (PIK), Potsdam 14473, Germany
}%
\affiliation{%
 Institute of Physics, Humboldt Universit\"at zu Berlin, Berlin 12489, Germany
}%

\author{R. I. Sujith}
\email{sujith@iitm.ac.in}
\affiliation{%
 Department of Aerospace Engineering, Indian Institute of Technology Madras, Chennai 600036, India
}%
\affiliation{%
 Centre of Excellence for Studying Critical Transitions in Complex Systems, Indian Institute of Technology Madras, Chennai 600036, India
}%

\date{\today}

\begin{abstract}

Self-sustained order can emerge in complex systems due to internal feedback between coupled subsystems. Here, we present our discovery of a non-monotonic emergence of order amidst chaos in a turbulent thermo-acoustic fluid system. Fluctuations play a vital role in determining the dynamical state and transitions in a system. In this work, we use complex networks to encode jumps in amplitude scales owing to fluctuations as links between nodes representing amplitude bins. The number of possible amplitude transitions at a fixed timescale reflects the complexity of dynamics at that timescale. The network entropy quantifies the number of and uncertainty associated with such transitions. Using network entropy, we show that the uncertainty in fluctuations first increases and then decreases as the system transitions from chaos via intermittency to order. The competition between turbulence and nonlinear interactions leads to such non-monotonic emergence of order amidst chaos in turbulent thermo-acoustic fluid systems.

\end{abstract}

\maketitle

\section{\label{sec:intro}Introduction}

Diverse systems exhibit the emergence of ordered dynamics from disorder. Fireflies flashing in synchrony  \cite{buck1976synchronous}, rhythmic applause of audience \cite{neda2000sound}, synchronous discharge of neurons causing epileptic seizures \cite{bromfield2006basic}, oscillatory chemical reactions in Belousov–Zhabotinsky reaction \cite{epstein1998introduction}, spatio-temporal patterns in surface chemical soup \cite{jakubith1990spatiotemporal}, biochemical oscillators such as glycolytic oscillators \cite{friesen1984biological}, pattern formation in fluid motion during convective instabilities \cite{berge1984rayleigh}, and the emergence of a single coherent electromagnetic wave in lasers \cite{oliva2001dynamics} are some typical examples of order emerging from disorder. Furthermore, the emergence of order in the form of oscillatory dynamics has been reported in turbulent thermo-acoustic fluid systems \cite{PhysRevE.89.022910,nair2016precursors,PhysRevApplied.7.044027,sujith2020complex,Sujith_2021}, and aerodynamic fluttering of slender structures such as bridges and wings of an aircraft \cite{jenkins2013self,venkatramani2016precursors}.

The emergence of ordered dynamics in such diverse systems is a consequence of self-organization and nonlinear interactions amongst the sub-units of the system \cite{bertalanffy1968general, kauffman1995home}. Additionally, the spatio-temporal structure of interactions, the functional behavior of sub-units, and the fluctuations in system variables are interdependent  \cite{prigogine1978time}. Fluctuations are inherently present in systems and are damped out for certain control parameter values. With a change in the control parameter, these fluctuations may amplify and aid in the selection of a new ordered dynamical state. Ordered dynamics, such as periodic oscillations, can be viewed as a giant macroscopic fluctuation in itself \cite{prigogine1978time}. 

The emergence of order through fluctuations, as first discussed by Prigogine \cite{prigogine1978time}, is an intriguing phenomenon and raises interesting questions about the occurrence of fluctuations. For example, what are the characteristic features of fluctuations at distinct timescales? How do these fluctuations evolve during the emergence of order? And, can we relate the variations in the amplitude scales of fluctuations to the dynamical transitions in the system? In this study, we analyze variations in fluctuations during the transition from chaos to order in the temporal dynamics of a complex thermo-acoustic fluid system. 

We perform experiments in a turbulent thermo-acoustic fluid system (Fig.~\ref{fig:ts_exp}(a)), which is known to exhibit dynamical transition from chaos to order. Thermo-acoustic systems find applications in gas turbine combustors and rocket engines. The turbulence intensity in a closed duct increases with increasing Reynolds number $Re$ \cite{avila2022transition}. {Turbulence introduces a wide range of time and length scales \cite{Frisch_1995,lesieur1987turbulence}, thus promoting chaotic dynamics in the acoustic field. As the inlet air flow rate increases, the system exhibits a transition from low-amplitude aperiodic to high-amplitude periodic fluctuations in the acoustic pressure signal. We note that even though the acoustic pressure exhibits self-sustained limit cycle oscillations, the underlying flow and flame dynamics are still turbulent. Such self-sustained oscillations emerge due to interactions between the flow, flame, and the acoustic subsystem \cite{chu_kovásznay_1958, lieuwen2005combustion, tandon_sujith_2023}.} 

Fluctuations in one subsystem depend on the function of other subsystems and the spatio-temporal structure of the inter-subsystem interactions \cite{tandon_sujith_2023}. While turbulence promotes disorder and chaos, nonlinear feedback between subsystems promotes self-organization and order. Although the range of spatial and temporal scales increases with $Re$, we observe the emergence of a single dominant frequency in the acoustic pressure signal, periodic shedding of large coherent structures, and sudden and intense heat release patterns occurring periodically \cite{george2018pattern}. Since the subsystems are co-evolving, the fluctuations in each subsystem determine as well as necessarily reflect the evolution of interactions between all subsystems. Thus, it is essential to study the role and characteristics of fluctuations during dynamical transitions.

{Turbulent thermo-acoustic fluid systems are known to exhibit a transition in the acoustic field from low-amplitude aperiodic fluctuations (which is high-dimensional chaos \cite{vineeth2013,tony2015}) as shown in Fig. 1(d), to high-amplitude periodic oscillations (order) as shown in Fig. 1(f) with an increase in the Reynolds number ($Re$) of the inlet flow \cite{lieuwen2005combustion,Sujith_2021,Pavithran_2020,Nair_Sujith_2014,sudarsanan_sivakumar_pre}.} Such a transition occurs via the route of intermittency \cite{nair2014intermittency}, which is characterized by bursts of high-amplitude periodic oscillations amidst epochs of low-amplitude chaotic oscillations in the acoustic field (Fig.~\ref{fig:ts_exp}(e)). The ordered dynamics observed at high value of $Re$ is referred as thermo-acoustic instability, characterized by large amplitude and a limit cycle attractor \cite{lieuwen2005combustion,Sujith_2021}.

Self-sustained periodic dynamics during thermo-acoustic instability has also been viewed as synchronization between two coupled non-identical oscillators representing the heat release rate and acoustic pressure fluctuations \cite{pawar2017thermoacoustic} in a combustor. Further, complex networks have been used to analyze the characteristics of temporal dynamics  \cite{okuno2015dynamics,murugesan2015combustion,godavarthi2017recurrence,gotoda2017characterization,doi:10.1063/1.5052210,kobayashi2019early,PhysRevE.99.032208,tandon2021condensation} during the transition from chaos to order in such systems. The presence of scale-invariant fluctuations during chaotic dynamics and the loss of such scale-invariance during the emergence of order was discovered by means of visibility graphs  \cite{murugesan2015combustion}. Recently, the transition from chaos to order in turbulent thermo-fluid systems was shown to be analogous to a transition from defect to phase turbulence in a system of diffusively coupled nonlinear oscillators \cite{García-Morales_2024}. Further, complex networks built from the phase space cycles delineate the transformation in the topology of the phase space from a set of several unstable periodic orbits to a single stable limit cycle orbit as the system dynamics transition from chaos via intermittency to order \cite{tandon2021condensation}.

Here, we use complex networks to encode the fluctuations in the temporal dynamics of ($p'$). A network is derived from time series of acoustic pressure fluctuations ($p'$) obtained from experiments in a turbulent combustor, where amplitude bins are nodes, and the weight of a link connecting two nodes defines the probability of transition between the respective amplitude bins \cite{shirazi2009mapping,yuan2023multi,ZOU20191}. We refer to the resulting network as amplitude transition network. We construct these networks at distinct timescales and examine the network properties. Using network entropy, we discover that the uncertainty and the number of amplitude transitions first increase and then decrease during the transition from chaos to order in the system, delineating a non-monotonic emergence of ordered dynamics from disorder. We also characterize the amplitude scales in the fluctuations during the emergence of order using distance-based centrality measures. 

The structure of the article is as follows: The details of the experimental setup and data acquisition are described in Sec.~\ref{exp_methods}. The network construction and the choice of parameters involved are explained in Sec.~\ref{network_const}. The measures derived from the network and their interpretations are described in Sec.~\ref{results}. Our conclusions are given in Sec.~\ref{conclusion}. {We have also confirmed that the experimental data presented in our study show that the transition to periodic dynamics indeed occurs via intermittency, as detailed in Appendix.~\ref{appendix_1}.}

\section{\label{methods} Methods}
\subsection{\label{exp_methods} Experimental setup}

\begin{figure*}
    \centering
    \includegraphics[width=\linewidth]{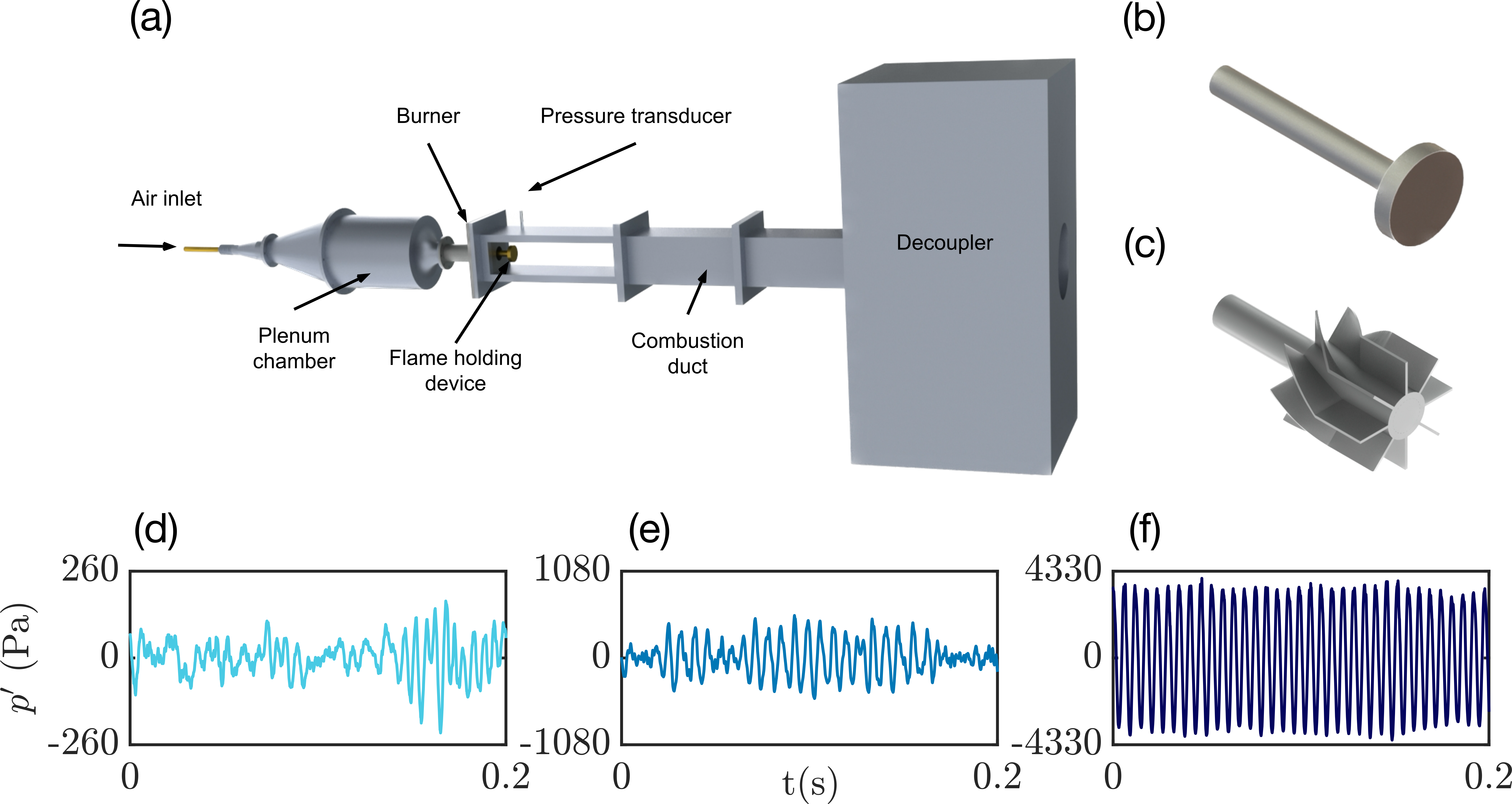}
    \caption{{The experimental setup and the time series of acoustic pressure fluctuations corresponding to different dynamical states.} (a) Schematic of the turbulent combustor used in this study. Air at room temperature enters the plenum chamber through the inlet and gets mixed with the fuel before entering the combustion chamber, where combustion occurs. The flame-holding device in the combustion duct is essential to maintain sustained combustion. The flame-holding devices that were used to perform experiments are -- (b) circular bluff-body and (c) vane swirler. The time series of acoustic pressure variation in a bluff-body stabilized turbulent combustor during the state of (d) low-amplitude aperiodic fluctuations (chaos) ($Re = 3.865 \times 10^{4}$) (e) intermittency ($Re = 5.083 \times 10^{4}$) (f) thermo-acoustic instability (order) ($Re = 6.759 \times 10^{4}$).} 
    \label{fig:ts_exp}
\end{figure*}

We conducted experiments in a turbulent combustor with two different types of flame-holding devices. A schematic of the turbulent combustor used in this study is shown in Fig.~\ref{fig:ts_exp}(a). The turbulent combustor comprises a plenum chamber, a burner, and a combustion chamber with an extension duct. Another chamber with a relatively larger cross-sectional area, known as the decoupler, is connected to one end of the combustion duct. The decoupler reduces the loss in acoustic energy through acoustic radiation. A central shaft of diameter 16 mm through the burner supports the flame-stabilizing circular disc object, i.e., a bluff-body (Fig.~\ref{fig:ts_exp}(b)) or a vane swirler (Fig.~\ref{fig:ts_exp}(c)).

{Compressed air enters the plenum chamber of the combustor before passing through the burner, where the mixing of the air and gaseous fuel (liquefied petroleum gas (LPG) -- 60\% butane and 40\% propane) takes place.} The central shaft delivers fuel into the combustion chamber via four radial injection holes with diameters of 1.7 mm each. The dynamics of the unsteady acoustic pressure oscillations in the combustor is investigated as the system parameter, i.e., the Reynolds number $Re$ of the incoming flow is varied. The acoustic pressure fluctuations in the combustion chamber are acquired using a piezoelectric pressure transducer (PCB103B02, with an uncertainty of $\pm$ 0.15 Pa). The transducer is mounted at a distance of 25 mm downstream from the dump plane and records the pressure signal for 3 s at a sampling rate of 10 kHz. 

{The $Re$ near the position of the burner is calculated as $Re = 4\dot{m}D_1/\pi \mu D_0^2$, where $\dot{m}$ is the sum of air flow rate and fuel flow rate ($\dot{m} = \dot{m_a} + \dot{m_b}$; in g/s), $D_0$ is the hydraulic diameter of the burner (in mm), $D_1$ is the diameter of the circular bluff-body (in mm; $D_1 = D_0$ for experiments with vane swirler) and $\mu$ is the dynamic viscosity of the fuel-air mixture at the experimental condition. We use the expression for dynamic viscosity $\mu$ as stated by Wilke \cite{wilke1950viscosity} to compute the $Re$ of a binary gas mixture.}

The fuel-air mixture is injected at turbulent conditions and undergoes mixing in recirculation zones created by a flame stabilizing mechanism. Here, we perform two sets of experiments using two types of flame stabilizers, namely, a bluff body and a vane swirler. 
The bluff body located at the end of the shaft forces the flow to turn near its shoulder, creating a large-scale recirculation zone downstream of the bluff body. On the other hand, a swirler is a non-rotating device with curved vanes (blade-like surfaces) placed radially around a central shaft, forcing the flow to turn along its vanes, creating a large-scale recirculation zone downstream of the swirler. These recirculation zones comprise hot radicals, and the large-scale recirculation vortices enhance the fuel-air mixing, which subsequently undergoes combustion. The heat released due to the combustion of the fuel-air mixture adds energy to the fluctuations in the acoustic field, leading to growth in their amplitude. Perturbations in the acoustic field influence the flame surface oscillations, leading to fluctuations in the heat release rate. Further, the acoustic and hydrodynamic fluctuations together influence the mixing and convection of the fuel-air mixture via coherent structures in the flow, which subsequently determine the dynamics of the turbulent thermo-fluid system \cite{chu_kovásznay_1958,lieuwen2005combustion}.

{For the experiments on the turbulent combustor with a bluff-body flame stabilizer, the fuel flow rate is maintained at a constant value at 1.76 g/s (grams per second), and the air flow rate is varied from 15.3 $-$ 29.5 g/s quasi-statically, in steps of 0.80 g/s. Hence, the corresponding $Re$ ranges from $(3.865 \pm  0.031) \times 10^4$ to $(6.759 \pm 0.054) \times 10^4$. Similarly, for the experiments on the turbulent combustor with swirl stabilized flame, we set a constant fuel flow rate of 0.76 g/s, and the air flow rate is varied from 6.73 $-$ 11.2 g/s quasi-statically, in steps of 0.20 g/s. Hence, the corresponding Reynolds number ranges from $(1.441 \pm 0.011) \times 10^4$ to $(2.220 \pm 0.018) \times 10^4$.}

\subsection{\label{network_const} Construction of amplitude transition networks}

We use the stochastic mapping network construction technique proposed by \textcite{shirazi2009mapping}, which explicitly encodes the temporal sequence of the considered variable. For a discrete time series $x(t)$ = $\{x_1, x_2, \ldots, x_n\}$, we begin the network construction by grouping these data points into horizontal amplitude-bins, i.e., discrete `states.' 
\begin{figure*}
    \centering
    \includegraphics[width=1\linewidth]{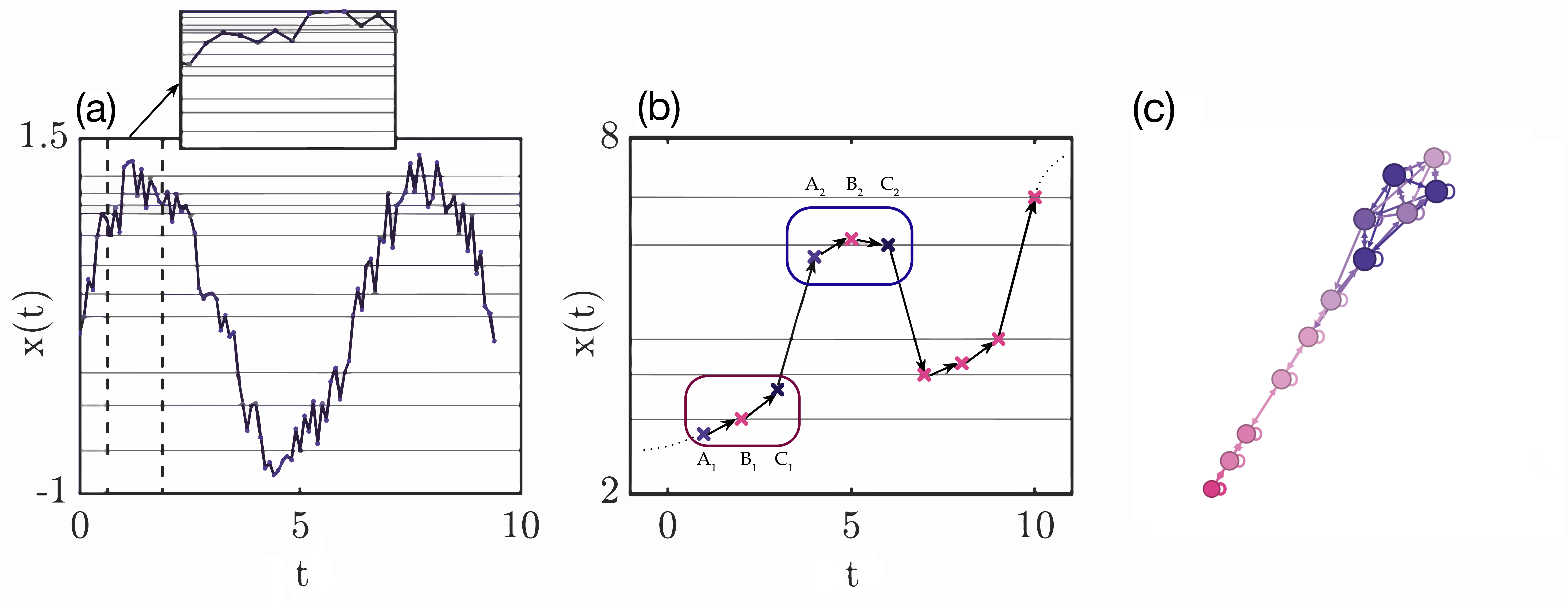}
    \caption{{Schematic for network construction.} (a) A noisy sinusoidal signal, $x(t) = \sin(x) + \xi$, where $\xi$ represents uniformly distributed random noise. There are 95 data points cumulatively partitioned into 12 bins, each containing approximately 8 data points. (b) Representative figure to examine the network formation for a transition timescale of $\tau$ = 3 time steps. (c) A stochastic mapping network constructed from the noisy sine signal. The constructed network is visualized with Force-Atlas algorithm \cite{jacomy2014forceatlas2} in the software Gephi \cite{Bastian_Heymann_Jacomy_2009}.}
    \label{fig:network_method}
\end{figure*}
The number of bins required and the respective width of each bin are chosen such that the number of data points in each bin is approximately equal. The number of bins ($N$) needed for this formulation is given by $N \approx 2n^{2/5}$ \cite{https://doi.org/10.18434/m32189}.

Each bin is considered a node in the network. There exists a link between node $i$ and $j$ if the variable $x(t)$ changes from $i^{th}$ bin to $j^{th}$ bin in one time step. We define the adjacency matrix $\textbf{A} = a_{ij}$, where $a_{ij}$ is the total number of transitions of the variable $x(t)$ from bin $i$  to $j$. We then compute the transitional probability matrix $\textbf{W} = w_{ij}$, where $w_{ij}$ is the probability of a Markov chain transition from node (bin) $i$ to $j$ and is given by $p[x(t) \mid x(t - 1)]$, i.e., $w_{ij} = {a_{ij}}/{\sum_k a_{ik}}$. We attribute $w_{ij}$ as the weight to the link from node $i$ to node $j$ in the network. Thus, $\textbf{W}$ is the weighted, asymmetric adjacency matrix of the network.

In our analyses, we modify the timescale of computing the transition probabilities of amplitude jumps by introducing a transition timescale $\tau$. This allows us to capture information from the network topology at different time scales. A transition matrix is constructed such that each entry is the transition probability $w_{ij} (\tau) = p[x(t) \mid x(t -\tau)]$, i.e., the probability of a jump from the $i^{th}$ to the $j^{th}$ amplitude bin after a time interval $\tau$. Figure~\ref{fig:network_method}(c) shows the network constructed from a noisy signal (Fig.~\ref{fig:network_method}(a)) with $\tau = 3$ time steps. Figure~\ref{fig:network_method}(b) shows a schematic series for counting transition when $\tau = 3$. We consider the transition between $A_1$ and $A_2$, $B_1$ and $B_2$, $C_1$ and $C_2$ in Fig.~\ref{fig:network_method}(b) and so on to find the number of transition from one bin to another. 

To facilitate comparison across different dynamical states, we consider a common set of nodes (amplitude bins) for all the time series of $p'$ obtained when $Re$ is varied in a quasi-static manner. The total number of bins for $p'$ obtained from experiments in bluff-body combustor is $N = 418$ and $N = 432$ in a swirl combustor.

\subsection{\label{} Measures from amplitude transition network}

Prigogine \cite{prigogine1978time} theorized that macroscopic order occurs in a system owing to the interplay between the function of the sub-units, the spatiotemporal structure of interactions, and the fluctuations in the system variables. The structure, function, and fluctuations in a system are interdependent, and a variation in either of them reflects the changes in the state of the system. In order to decipher the dynamical evolution of a turbulent thermo-acoustic fluid system, we characterize the fluctuations in the temporal dynamics of the acoustic subsystem by encoding the fluctuations into the topology of a complex network. 

{We construct amplitude transition networks using the normalized time series of acoustic pressure fluctuations ($p'$). The normalization is done with the maximum value of the acoustic pressure during the state of thermo-acoustic instability, specifically the maximum value of $p'$ obtained at the control parameter ($Re = (6.759 \pm 0.054) \times 10^4$ for experiments in bluff-body combustor and $Re = (2.220 \pm 0.018) \times 10^4$ for experiments in swirl combustor).} As described in Sec.~\ref{network_const}, a weighted and directed network is constructed for each dynamical state at different time scales $\tau$. The topology of the constructed network encodes the temporal order of fluctuations in the dynamics. Periodic oscillations would result in a regular network topology, where connections exist in orderly fashion between specific amplitude bins. On the other hand, chaotic fluctuations would be encoded as connections between nodes corresponding to non-adjacent amplitude bins and corresponding to different amplitude scales, and the network structure becomes more disorderly.

The Shannon entropy of the network ($\mu_S$) quantifies the network topology as defined by Eq.~\ref{eq:avg_entropy}  \cite{shannon1948mathematical,small2013complex,wiedermann2017mapping}: 

\begin{equation}
    \mu_S = - \frac{\sum_{j}\sum_{i} w_{ij} \ln{w_{ij}}}{N_t}
    \label{eq:avg_entropy}
\end{equation}

\noindent Here, $w_{ij}$ is the probability of the transition from node $i$ to $j$, and $N_t$ is the number of nodes in the network. For a complex network, the average network entropy quantifies the orderliness of connections between nodes and will be high when the network structure is more disordered \cite{wiedermann2017mapping}. For an amplitude transition network, the average Shannon entropy ($\mu_S$) reflects the complexity as it quantifies the number of possible transitions and uncertainty in the transition between nodes corresponding to different amplitude scales owing to fluctuations in the dynamics. Note that several measures of complexity have been proposed \cite{ladyman2013complex,shiner1999simple}, all of which primarily aim to quantify the variations in a number of possible states of a system across timescales \cite{estrada2023complex,ladyman2013complex}.

In amplitude transition network, at a specific timescale, the number of possible transitions from one amplitude scale to another is infinitely many. Yet, only a selected multitude of these transitions are realized during a certain dynamical state, reflecting the `complexity' of the fluctuation dynamics at the selected timescale. Complexity here simply refers to the number of possible realizations or transitions a variable can exhibit at a fixed timescale \cite{estrada2023complex,siegenfeld2020introduction}. For example, a periodic signal has the least complexity as only certain amplitude transitions are realized in an orderly manner in the system. Whereas, during chaos, many amplitude transitions are realized, and the occurrence of any one of these transitions is highly uncertain. Hence, we emphasize that entropy is a measure of complexity only at the timescale at which the amplitude transitions are encoded in the network \cite{estrada2023complex}. Also, Shannon entropy is equivalent to the Clausius entropy, which characterizes the macroscopic thermodynamic property of a system  \cite{weilenmann2016axiomatic}. Thus, $\mu_S$ of an amplitude transition network encodes the uncertainty in temporal fluctuations of a dynamical system and also quantifies the thermodynamic disorder of the system.

Furthermore, we explore the diversity in the magnitude of jumps between different amplitude scales associated with fluctuations. To do so, we use path-based centrality measures derived from amplitude transition networks. A path from one node to another in the network denotes the jump from one amplitude bin to another in the time series. Thus, we compute distance-based metrics, namely the characteristic path length (CPL) and mean betweenness centrality ($\overline{C}_{BC}$). 

The CPL of a network is the mean number of steps needed to reach one node from another in the network  \cite{barabasi_2002,RUBINOV20101059}. For amplitude transition networks, CPL encodes the magnitude of jump in amplitude scale due to fluctuations between different amplitude bins (nodes). For example, in a periodic time series, any amplitude bin (node) is predominantly connected to specific other amplitude nodes in an orderly manner. The differences in amplitude scales are small and of similar magnitude when the transition occurs from any one bin to its neighboring bin, and this is reflected as a high value of CPL. On the other hand, a small value of CPL indicates that fluctuations occur between nodes of very different amplitude scales, such as during chaotic dynamics.

The betweenness centrality of a node quantifies the extent to which a node lies in the path between other nodes of the network  \cite{freeman1977set,RUBINOV20101059}. For an amplitude transition network, the mean betweenness centrality $\overline{C}_{BC}$ quantifies the order in which transition occurs between different amplitude scales. If there are direct transitions between bins of diverse amplitude scales, such as during chaotic fluctuations, then the value of $\overline{C}_{BC}$ will be low. However, for a periodic time series, the connections are mostly between specific amplitude nodes, and any node lies `in between' the path of many other node pairs with different amplitude scales, resulting in a high value for $\overline{C}_{BC}$.

\section{\label{results} Results}

\begin{figure*}
    \centering
    \includegraphics[width=\linewidth]{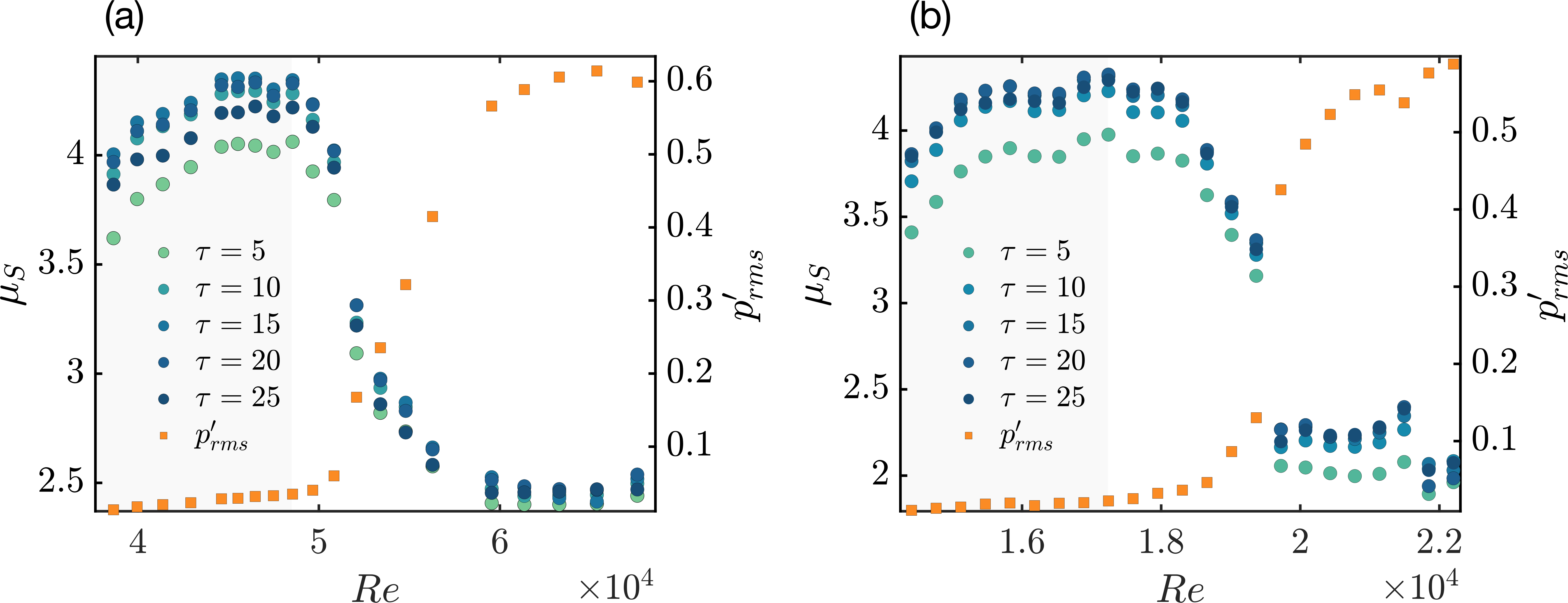}
    \caption{{The average Shannon entropy derived from amplitude transition networks, which quantifies the amount of disorder, first increases and then decreases during the transition from chaos to order for different transition time-scales $\tau$ (in $10^{-4}$ s). (a) Variation of average network entropy ($\mu_S$) and root-mean-square of the non-dimensional pressure fluctuations $p'_{rms}$ with the flow control parameter $Re$ for a bluff-body stabilized turbulent combustor. The shaded region corresponds to the region where $Re < (Re_c)_{\text{bluff}}$ where $(Re_c)_{\text{bluff}} = 4.966 \times 10^4$. (b) Variation of average network entropy ($\mu_S$) and root-mean-square of the pressure fluctuations ($p'_{rms}$) with the flow control parameter $Re$ for a swirl stabilized turbulent combustor. The shaded region corresponds to the region where $Re < (Re_c)_{\text{swirl}}$ where $(Re_c)_{\text{swirl}} = 1.724 \times 10^4$.}}
    \label{fig:diff_tau_entropy_bluff_swirl}
\end{figure*}

Figure~\ref{fig:diff_tau_entropy_bluff_swirl} shows the variation in $\mu_S$ of networks derived at different transition time scales $\tau$ (in $10^{-4}$ s; as the sampling frequency of the time series is 10 kHz) from the acoustic pressure signal as the control parameter ($Re$) is varied quasi-statically in experiments. We find that the average network entropy $\mu_S$ varies non-monotonically with $Re$ during the transition from chaos to order in both bluff-body (Fig.~\ref{fig:diff_tau_entropy_bluff_swirl}(a)) and swirl stabilized combustors (Fig.~\ref{fig:diff_tau_entropy_bluff_swirl}(b)) in a similar manner for several transition time-scales $\tau$. {We define the critical Reynolds number $Re_c$ as the control parameter $Re$ for which the average network entropy $\mu_S$ obtained from the time series of $p'$ attains a maximum value. For the bluff-body stabilized combustor, the \(Re_c\) is $(4.966 \pm 0.040) \times 10^4 $, whereas for the swirl combustor, it is $(1.724 \pm 0.014) \times 10^{4}$.} In Appendix.~\ref{appendix_2}, we show that the trend in network entropy is largely unaffected with variation in the other network construction parameter, i.e., the number of amplitude bins ($N$).

\begin{figure*}
    \centering
    \includegraphics[width=\linewidth]{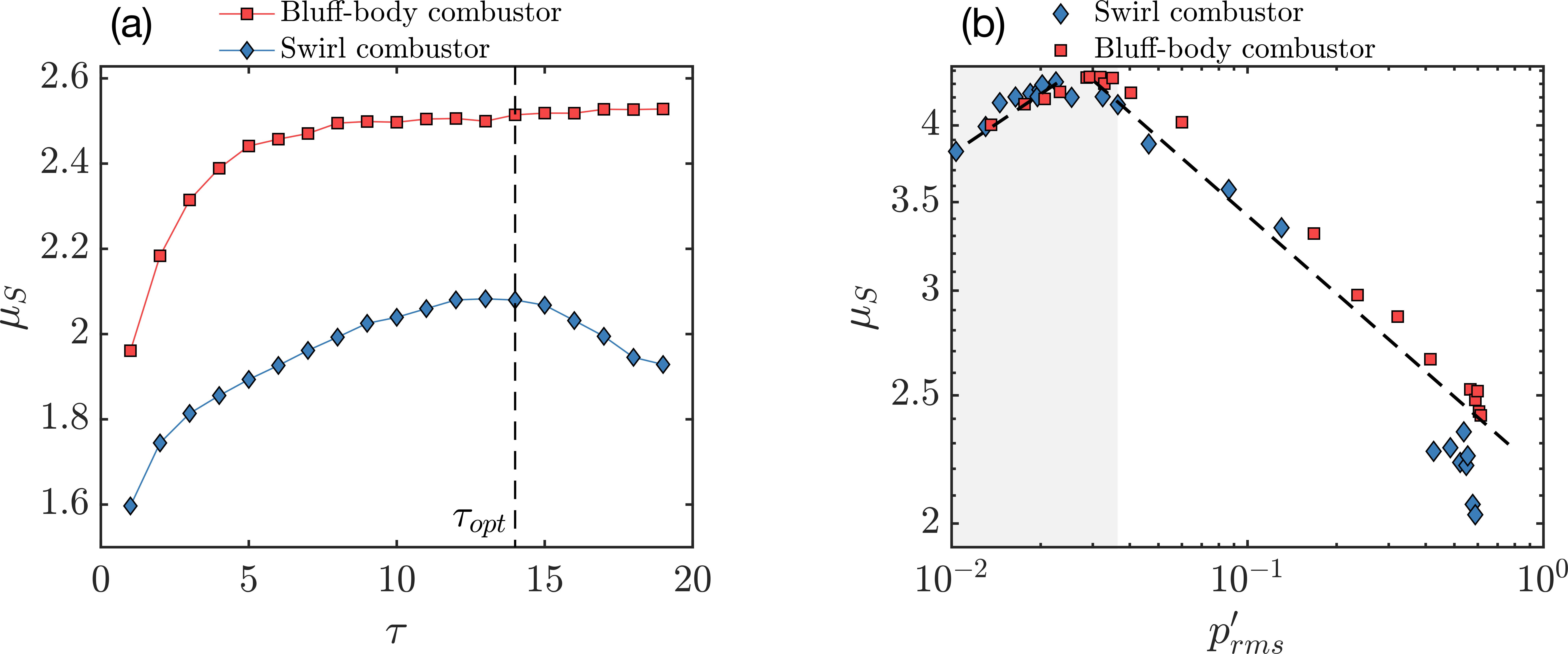}
    \caption{(a) Variation of average Shannon entropy ($\mu_S$) of the network constructed from the time series corresponding to the dominant acoustic mode at different transition time scales $\tau$ (in $10^{-4}$ s). We chose the optimal time-scale to be one that maximizes the information obtained from the network, i.e., $\tau_{opt} = 14 \times 10^{-4}$ s for both bluff-body and swirl stabilized turbulent combustors. (b) $\mu_S$ from both bluff-body and swirl combustor corresponding to $\tau_{opt}$ exhibits a power law scaling $(\mu_S)_{\text{bluff}} \propto (p'_{rms})^{\alpha_{i}}$, and $(\mu_S)_{\text{swirl}} \propto (p'_{rms})^{\beta_{j}}$, respectively with similar exponents ($i $ and $j$ $\in$  \{1,2\}). The power law exponents for a bluff-body combustor when $Re < (Re_c)_{\text{bluff}}$ is $\alpha_1 = 0.09 \pm 0.025$, and when $Re > (Re_c)_{\text{bluff}}$ is $\alpha_2 = -0.20 \pm 0.009$. Similarly, the power law exponents for a swirl combustor when $Re < (Re_c)_{\text{swirl}}$ is $\beta_1 = 0.15 \pm 0.049$, and when $Re > (Re_c)_{\text{swirl}}$ is $\beta_2 = -0.21 \pm 0.020$.}
    \label{fig:bluff_swirl_entropy}
\end{figure*}

\begin{figure*}
    \centering
    \includegraphics[width=\linewidth]{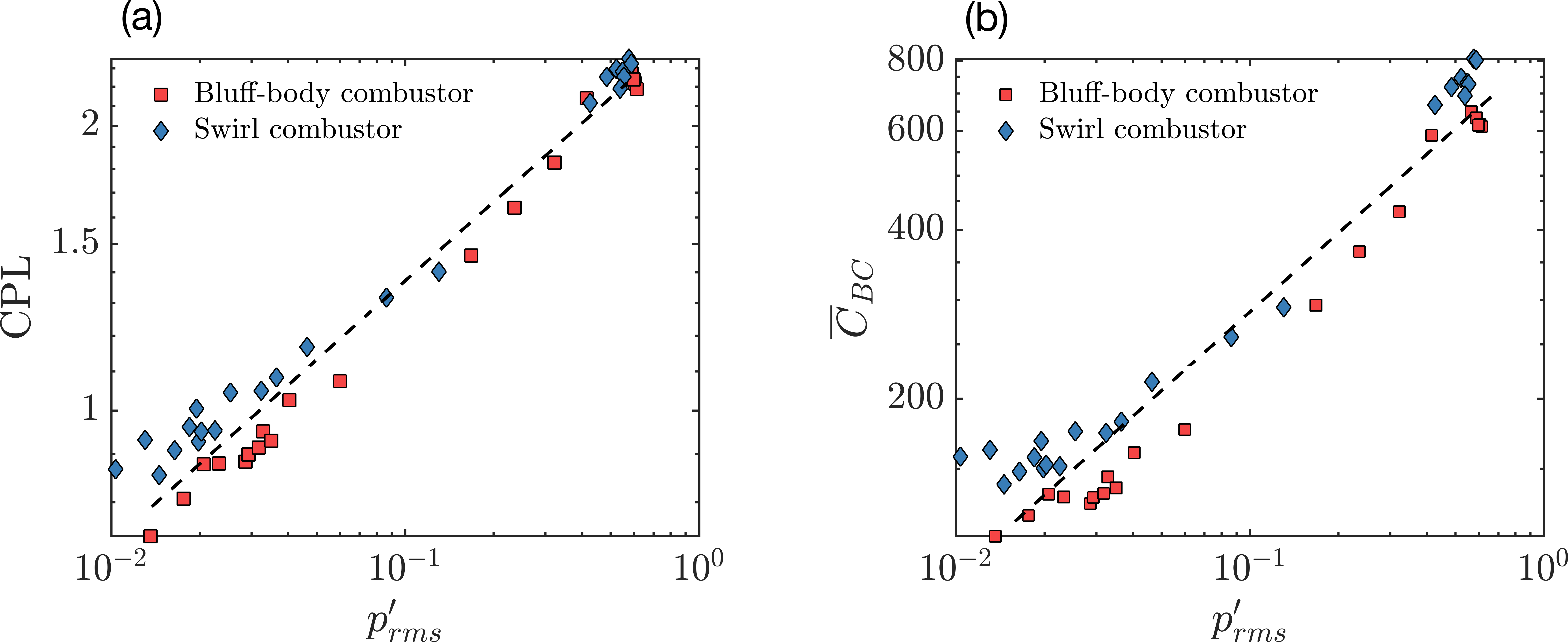}
    \caption{Quantifying the diversity in amplitude scale of fluctuations using distance-based network topological metrics. (a) Variation of characteristic path length (CPL) of the network constructed from time series of non-dimensional $p'$ of bluff-body and swirl stabilized turbulent combustor with $p'_{rms}$ (i.e., for the entire range of control parameter $Re$) at $\tau_{opt}$. The measure from both combustors exhibits the same power law scaling $\text{CPL} \propto (p'_{rms})^{\gamma}$, with exponent $\gamma = 0.28 \pm 0.016$. (b) Variation of average betweenness centrality ($\overline{C}_{BC}$) of network constructed from time series of non-dimensional $p'$ of bluff-body and swirl stabilized turbulent combustor with $p'_{rms}$ at $\tau_{opt}$. $\overline{C}_{BC}$ from the bluff-body and swirl combustor exhibits a similar power law scaling $\overline{C}_{BC} \propto (p'_{rms})^{\kappa}$ with exponents $\kappa = 0.48 \pm 0.033$ and $\kappa = 0.47 \pm 0.036$, respectively.}
    \label{fig:centrality}
\end{figure*}

Clearly, from Fig.~\ref{fig:diff_tau_entropy_bluff_swirl}, at each transition time scale, the number of possible amplitude transitions and the respective uncertainties in $p'$ and hence the complexity of the dynamics first increase and then decrease with $Re$. This implies that during the transition from chaos via intermittency to order, the complexity of the dynamics is maximum somewhere between chaos and order in our system. This finding corroborates the most accepted description of maximum complexity between complete disorder and order \cite{wackerbauer1994comparative, shiner1999simple}. 

{Next, we chose an optimal transition timescale ($\tau_{opt}$) as the value of $\tau$ for which $\mu_S$ of the amplitude transition network of $p'$ during the state of limit cycle oscillations is maximum and/or saturates. We obtain $\tau_{opt} = 14 \times 10^{-4}$ s for both the bluff-body and the swirl combustor (refer Fig. 4(a)), which is approximately $T/4$ s and $T/6$ s, respectively (where $T$ is the time period of $p'$ during thermo-acoustic instability in each combustor configuration).} The variation of $\mu_S$ with $p'_{rms}$ at $\tau_{opt}$ is shown in Fig.~\ref{fig:bluff_swirl_entropy}(b). We find that $\mu_S$ exhibits a power law scaling with $p'_{rms}$ and the power law exponents are similar for the results obtained from both set of experiments (bluff-body and swirler configuration). Such scaling relations are of particular interest as it helps relate the uncertainty in amplitude transitions to the amplitude itself. Evidently, this relation is similar for different combustors and indicates universality in the features of the dynamical transition rather than dependence on the specific configuration of the system.

{The non-monotonicity of entropy during the emergence of order reveals that during the transition from chaotic to periodic dynamics, there is an interplay between the mechanisms that promote order and disorder. A thermo-acoustic system is a complex system where the acoustic, combustion, and hydrodynamic subsystems are nonlinearly coupled. Positive feedback between these subsystems and spatially organized interactions can lead to synchronized co-evolution of variables in these subsystems, promoting pattern formation and ordered dynamics \cite{tandon_sujith_2023,george_unni_raghunathan_sujith_2018}. On the other hand, as we increase the Reynolds number of the inlet flow, the turbulent intensity in the confined duct increases, promoting disorder and aperiodic fluctuations in the acoustic pressure. Note that both the mechanisms promoting aperiodic and periodic fluctuations in $p'$ occur throughout all dynamical states during the transition from chaos to order. Our findings imply that disorder created by increasing turbulence intensity is prominent until a critical control parameter $Re_c$. Beyond $Re_c$, the nature of interactions between subsystems changes, perhaps becoming more spatially organized \cite{tandon_sujith_2023}, and hence the periodicity is sustained for longer epochs. As a result, disorder decays monotonically despite the increase in $Re$ beyond $Re_c$. Moreover, the decrease in $\mu_S$ for $Re>Re_c$ is a precursor for an impending thermo-acoustic instability.}

Further, the variation of CPL and $\overline{C}_{BC}$ with $p'_{rms}$ derived from a bluff-body and swirl-stabilized turbulent combustor is shown in Fig.~\ref{fig:centrality}(a) and Fig.~\ref{fig:centrality}(b), respectively. Clearly, both measures are minimum during chaotic dynamics and reach a maximum during periodic dynamics (thermo-acoustic instability). We infer that the variation in the magnitudes of jumps in amplitude scales due to fluctuations decreases in a monotonic fashion as order emerges amidst chaos in the system. Intriguingly, during the transition from chaos to order with increase in $Re$, we discover a regime ($Re < Re_c$) where fluctuations become more disordered and uncertain, and hence, the network entropy ($\mu_S$) increases, while the amplitude jumps associated with the fluctuations become less diverse. For $Re > Re_c$, both the number of possible transitions and the range of amplitude scales in which the fluctuations occur decrease. Moreover, CPL and $\overline{C}_{BC}$ exhibit distinct power-law scaling behavior with the root-mean-square of acoustic pressure fluctuations. The power law exponents are nearly identical for experiments in both bluff-body and swirl-stabilized turbulent combustors. 

In summary, we used complex networks to characterize the fluctuations and the jumps in amplitudes due to fluctuations as a dynamical transition occurs in a complex turbulent thermo-fluid system.

\section{\label{conclusion} Discussion and Conclusions}

Several complex systems exhibit self-organization owing to an ensemble of elements that interact nonlinearly with each other. The nonlinear interactions between sub-units of a complex system can lead to continuous or abrupt emergence of self-sustained ordered dynamics. For example, a continuous transition to ordered dynamics occurs in a system of chemical oscillators \cite{kiss2002emerging}, whereas order emerges abruptly in the magnetic properties of ferromagnetic materials \cite{kouvel1961abrupt}. Here, we discover that order emerges non-monotonically in a turbulent reacting flow confined in a duct coupled to the acoustic field of the duct.

We study the dynamical transition from chaos to order in a turbulent thermo-fluid system by analyzing the fluctuations in one of the subsystems. Fluctuations in system variables reflect the interactions between underlying sub-processes. At the same time, these fluctuations also affect the functional form of the sub-units of a complex system. We use a complex network approach to encode the fluctuations in the system dynamics and apply network measures to quantify the characteristics of fluctuations during the transition from chaos to order in the system. The network is constructed by binning amplitudes and encoding the transition probabilities due to fluctuations between these bins and is referred to as amplitude transition networks. The average entropy $\mu_S$ of an amplitude transition network captures the uncertainty in the amount of fluctuations between different amplitude scales and reflects the complexity of the dynamics at a fixed time scale. On the other hand, distance-based network metrics (CPL and $\overline{C}_{BC}$) indicate the diversity in the jumps in amplitude scales due to fluctuations in the time series. Thus, our study offers a novel outlook in characterizing fluctuations, which plays a crucial role in the emergent of ordered dynamics \cite{prigogine1978time}.

We find that, during the transition from chaos to order in a complex turbulent thermo-fluid system, $\mu_S$ first increases and then decreases, whereas CPL and $\overline{C}_{BC}$ increase monotonically. Thus, the uncertainty of transition between amplitude bins decreases non-monotonically, whereas the diversity of magnitude of amplitude jumps decreases monotonically during the emergence of order from chaos in a turbulent thermo-acoustic fluid system. Also, the network entropy can serve as an effective precursor for the impending instability as the measure decreases much before the onset of thermo-acoustic instability. 

Moreover, we observed that the scaling behavior for all the measures is similar for two different experimental configurations using different flame-holding devices in a turbulent combustor. These findings point to the similarities in these turbulent combustors, although they have quite different underlying flame and flow physics. Our findings imply that the underlying turbulence and the nonlinear feedback between subsystems compete to dominate the dynamics of the system. 

{We identify a critical value $Re_c$, such that for $Re<Re_c$, disorder increases with $Re$ in the form of uncertainty in fluctuations in the acoustic pressure. However, beyond $Re_c$, disorder decreases despite the increase in $Re$ owing to strong positive feedback and organized interactions between subsystems. Self-organization becomes prominent beyond $Re_c$ leading to the emergence of self-sustained ordered dynamics in the acoustic field despite the turbulence in the underlying flow and flame dynamics.} The role of fluctuations in a complex system exhibiting emergent dynamics can be analyzed for various natural and engineering systems using the approach presented in this work. By encoding the fluctuations as the topology of a network, we also encode the structure of interactions in the system that influence these fluctuations. Additionally, we propose that using measures derived from amplitude transition networks, one can quantify as well as compare the different types of dynamical transitions across systems, such as monotonic, non-monotonic, or abrupt transitions between chaos and order. 

{An analytical characterization of the non-monotonic transition would explain several features of the dynamical transition in turbulent thermo-acoustic systems would be an interesting challenge for future work. Recently, the transition to periodic oscillations was modeled as a transition from defect to phase turbulence \cite{García-Morales_2024}. This was achieved using several Stuart Landau oscillators that are both diffusively and globally coupled through a globally coupled Complex-Ginzburg Landau equation, capturing both qualitative and quantitative features of the system. A possible direction would be extending this model to study the relation between amplitude fluctuations and dynamical transition in the acoustic field.} \\~\\

\noindent \textbf{Data availability:} The data used in this study are available from the corresponding author upon reasonable request.

\begin{acknowledgments}
The authors thank S. Sivakumar, S. Thilagaraj, S. Anand (Indian Institute of Technology, Madras), P. R. Midhun (Aalto University, Finland), Anaswara Bhaskaran (University of Pavia, Italy) and Dr. Samadhan A. Pawar (University of Toronto) for providing the experimental data. This research was supported by J. C. Bose Fellowship (JCB/2018/000034/SSC) and the IoE-IITM Research Initatives (SP/22-23/1222/CPETWOCTSHOC) grants provided to R.I.S.. S.T. acknowledges the support from Prime Minister's Research Fellowship, Govt. of India. The authors also acknowledge Mr. Radhakrishnan and Dr. Raghunathan for their valuable discussions. \\~\\
\end{acknowledgments}

\noindent \textbf{Author contributions:} Conceptualization: S.T., R.I.S. Methodology and analysis: A.B., S.T. Interpretation: A.B., S.T., N.M., J.K., R.I.S. Visualization: A.B. Supervision: N.M., J.K., R.I.S. Writing—original draft: AB, ST. Writing—review \& editing: A.B., S.T., N.M., J.K., R.I.S.

\appendix

\section{{Characterizing intermittency dynamics using recurrence plots}} \label{appendix_1}

{The transition from chaos to ordered dynamics via the route of intermittency in turbulent thermo-acoustic systems was reported first by Nair \textit{et al.} \cite{nair2014intermittency} and subsequently by others \cite{steinberg_inter, SAMPATH2016309, KHEIRKHAH2017319, aoki2020, Guan_Gupta_Li_2020, BONCIOLINI2021396, Hiroshi_2016}.}

{Recurrence plots help to visualize and quantify the architectural features of a time series \cite{marwan_2007}. These plots encode the time taken for a trajectory to recur at a location in the phase space. The recurrence matrix has entries of ones (1s) corresponding to the time stamps when the dynamics recur in the phase space and zeros otherwise. The patterns in the recurrence matrix characterize the dynamical state of the system. For example, parallel diagonal lines signify the periodic nature of the time series, and the distance between the parallel lines denote the time period of oscillations. }
\begin{figure*}
    \centering
    \includegraphics[width=\linewidth]{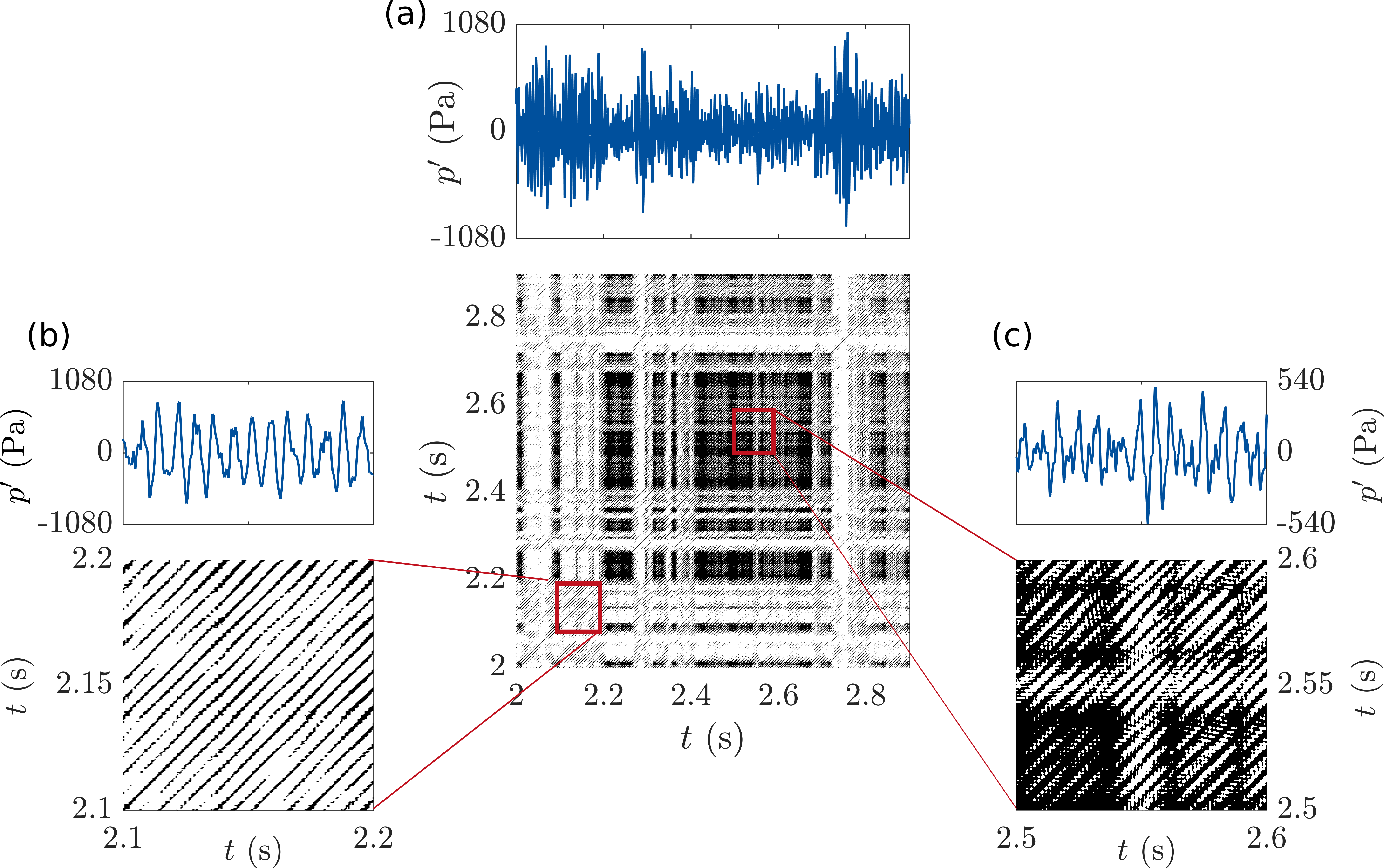}
    \caption{{(a) Time series of acoustic pressure during intermittency ($Re = (4.854 \pm 0.039) \times 10^4$) and its corresponding recurrence matrix. The insets (b) and (c) shows the topology of recurrence matrix during periodic and aperiodic fluctuation epochs of the pressure signal, respectively.}} 
    \label{fig:recurrence}
\end{figure*}
{Recurrence plots can help to identify and distinguish intermittency dynamics in time series data \cite{inter_recurrence}. A time series is embedded in a phase space using Takens' delay embedding theorem \cite{takens1981lecture}. According to Takens' theorem, we construct the phase space using the delayed time series $x(t+n\tau)$, where $n\in[0,m]$ and $m$ is the embedding dimension. Also, $\tau$ is the time lag for which the time series $x(t)$ and $x(t+\tau)$ have minimum mutual information. Next, a recurrence plot is constructed such that both the abscissa and ordinate represent time stamps. If the phase space trajectory visits the same location (within a reasonable bound $\varepsilon$) at two different time stamps $t_i$ and $t_j$, then the entry in the recurrence plot at $(t_i,t_j)$ and $(t_j,t_i)$ is set to value one indicating recurrence of the trajectory in the phase space. Further details for phase space construction and recurrence plot construction are presented in \textcite{nair2014intermittency}. }

{We find the embedding dimension and the lag for delay embedding for phase space reconstruction is 6 and 1.6 ms, respectively. Next, we construct a recurrence matrix from the time series of acoustic pressure $p'$ during the state of intermittency at the control parameter $Re = (4.854 \pm 0.039) \times 10^4$ shown in Fig.~\ref{fig:recurrence}(a). We have performed recurrence analysis for a window of $0.8$ s of the time series to show the patterns in the RP clearly. In our analysis, the trajectory recurs if the distance between the two points in the phase space is less than $\varepsilon$, where $\varepsilon$ is approximately 1/5$^{\text{th}}$ of the size of the attractor \cite{nair2014intermittency}. }

{The recurrence matrix shown in Fig.~\ref{fig:recurrence} has regions of parallel diagonal lines (Fig.~\ref{fig:recurrence}(b)) interspersed between black patches (Fig.~\ref{fig:recurrence}(c)). This topology of the recurrence matrix depicts the intermittent nature of acoustic pressure signal as there is a co-existence of black patches signifying aperiodicity and diagonal lines signifying periodicity \cite{inter_recurrence}. The black patches correspond to low-amplitude aperiodic fluctuations (see $t \in [2.5, 2.6]$; Fig.~\ref{fig:recurrence}(c)) and the diagonal lines represent the high-amplitude periodic oscillations (see $t \in [2.1, 2.2]$; Fig.~\ref{fig:recurrence}(c)).}

\section{Effect on the number of amplitude bin used in network construction} \label{appendix_2}

\begin{figure*}
    \centering
    \includegraphics[width = \linewidth]{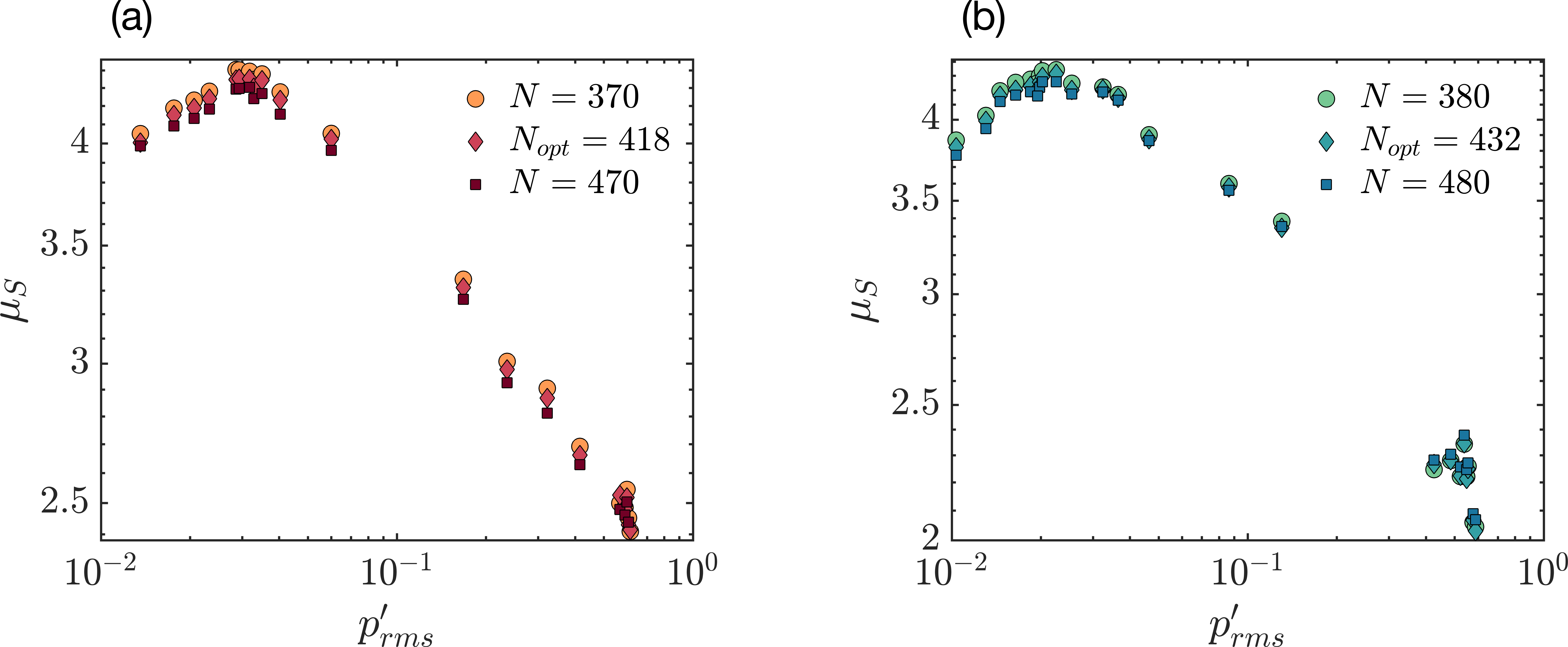}
    \caption{{Variation of $\mu_S$ with $p'_{rms}$ for different amplitude bins used for network construction from (a) bluff-body and (b) swirl stabilized turbulent combustor (at $\tau_{opt}$). The trend in $\mu_S$ is robust to different bin counts near the optimal value (obtained as discussed in the methods section) for both experimental configurations.}}
    \label{fig:appendix_2}
\end{figure*}
In this section, we verify the robustness of the results presented in this article with variation in the number of amplitude bins ($N$) considered for network construction. To do so, we vary the bin count from the optimal value discussed in Sec.~\ref{network_const} and see the effect on $\mu_S$ variation for both bluff-body and swirl stabilized turbulent thermo-acoustic systems (Fig.~\ref{fig:appendix_2}(a) and Fig.~\ref{fig:appendix_2}(b) respectively). It is evident from Fig.~\ref{fig:appendix_2} that the non-monotonous trend of $\mu_S$ is retained, and we have also verified that the scaling exponents are largely unaffected.

\bibliography{ref}

\end{document}